\documentclass{article}
\usepackage{spconf}
\usepackage{algorithm,graphicx}
\usepackage{algorithmic}
\usepackage{url}
\usepackage{cite}
\usepackage[usenames]{color}
\usepackage{amsfonts}
\usepackage{fancyhdr}
\usepackage{listings}%
\usepackage{arydshln}
\usepackage{amsmath,amssymb,amsthm}%
\usepackage{graphicx,psfrag,caption,subcaption}
\usepackage{listings}%
\usepackage{arydshln}
\usepackage{algorithmic}%
\input{mysymbol.sty}

\title{Learning Product Graphs from Multidomain Signals}
\name{Sai Kiran Kadambari and  Sundeep Prabhakar Chepuri 
}
\address{Indian Institute of Science, Bangalore, India
}

\begin{document}
\ninept
\maketitle

\begin{abstract}
In this paper, we focus on learning the underlying product graph structure from multidomain training data. We assume that the product graph is formed from a Cartesian graph product of two smaller factor graphs. We then pose the product graph learning problem as the factor graph Laplacian matrix estimation problem. To estimate the factor graph Laplacian matrices, we assume that the data is smooth with respect to the underlying product graph.  When the training data is noise free or complete, learning factor graphs can be formulated as a convex optimization problem, which has an explicit solution based on the water-filling algorithm. The developed framework is illustrated using numerical experiments on synthetic data as well as real data related to air quality monitoring in India.
\end{abstract}

\begin{keywords}
Graph learning, graph signal processing, Laplacian matrix estimation, product graphs, task-cognizant graph learning, topology inference.
\end{keywords}
\vspace*{-1mm}
\maketitle

\section{Introduction}
Leveraging the underlying structure in data is central to many machine learning and signal processing tasks~\cite{Bigdata,shuman2012emerging,smola2003kernels,cai2010graph}. In many cases, data resides on irregular (non-Euclidean) domains. Some examples include datasets from  meteorological stations, traffic networks, social and biological networks, to name a few. 

Graphs offer a natural way to describe the structure and explain complex relationships in such network datasets. More specifically, data (signal) is indexed by the nodes of a graph and the edges encode the relationship between the function values at the nodes. When the number of nodes in the graph is very large, signal processing or machine learning operations defined on graphs require more memory and computational resources, e.g., computing the graph Fourier transform~\cite{shuman2012emerging} using an eigendecomposition requires $O(N^3)$ operations for a graph with $N$ nodes~\cite{CooleyTukey}. Whenever the underlying graph can be factorized into two or more factor graphs with fewer nodes, the computational costs of these operations can be reduced significantly~\cite{Bigdata}. Product graphs can efficiently represent multidomain data. For instance, in brain imaging, the data is multidomain as each spatially distributed sensor gathers temporal data. Similarly, in movie recommender systems (such as Netflix) the rating data matrix has a user dimension as well as a movie dimension. The graph underlying such multidomain data can be often be factorized so that each graph factor corresponds to one of the domains. 

Having a good quality graph is essential for most of the machine learning or signal processing problems over graphs. However, in some applications, the underlying graph may not readily available, and it has to be estimated from the available training data. Given the training data, the problem of estimating the graph Laplacian or the weighted adjacency matrix, assuming that the graph signals are smooth on the underlying topology has been considered in\cite{learnDong,kalofolias2016learn,chepuri2016learning,kalofolias2017learning,LearnGraphData,egilmez2017graph}.  Even though the available training data might be multidomain, existing graph learning methods ignore the product structure in the graph. For example, when dealing with data collected from spatially distributed sensors over a period of time, existing methods learn a graph that best explains the spatial domain while ignoring the temporal structure. In \cite{kalofolias2017learning},  time-varying graphs are estimated from smooth signals, where the second domain is assumed to be regular.

In this paper, instead of ignoring the structure in any of the domains or treating one of the domains as regular, we propose to learn the underlying graphs related to each domain. This corresponds to learning the factors of the product graph. Concretely, the contributions of this paper are as follows.
We propose a framework for estimating the graph Laplacian matrices of the factors of the product graph from the training data. The product graph learning problem can be solved optimally using a {\it water-filling} approach. Numerical experiments based on synthetic and real data related to air quality monitoring are provided to demonstrate the developed theory.  Since the real dataset has many missing entries, we present an alternating minimization algorithm for {\it joint matrix completion and product graph learning}. 


Throughout this paper, we will use upper and lower case boldface letters to denote matrices and column vectors, respectively. We will denote sets using calligraphic letters. ${\bf 1}$ (${\bf 0}$) denotes the vector/matrix of all ones (zeros) of appropriate dimension. $\bbI_P $ denotes the identity matrix of dimension $P$. $\diag[\cdot]$ is a diagonal matrix with its argument along the main diagonal. 
${\rm vec}(\cdot)$ denotes the matrix vectorization operation. $X_{ij}$ and 
$x_i$ denote the $(i,j)$th element and $i$th element of $\bbX$ and $\bbx$, respectively. $\oplus$ represents the Kronecker sum and $ \otimes $ represents the Kronecker product.

\section{Product graph signals} \label{sec:productgraphs}

Consider a graph $\ccalG_{N} = (\ccalV_{N},\ccalE_{N})$ with $N$ vertices (or nodes), where $\ccalV_{N}$  denotes the set of vertices
and $\ccalE_{N}$ denotes the edge set. The structure of the graph with $N$ nodes is captured by the weighted adjacency matrix $\bbW \in \reals^{N \times N} $ whose $(i,j)$th entry denotes the weight of the edge between node $i$ and node $j$. When there is a no edge between node $i$ and node $j$, the $(i,j)$th entry of $\bbW$ is zero. We assume that the graph is undirected with positive edge weights. The corresponding graph Laplacian matrix is a symmetric matrix of size $ N $, given by  $ \bbL_{N} = {\diag}[\bbd] -\bbW$, where $\bbd \in \reals^N$ is a degree vector given by $ \bbd = \bbW{\bf 1} $.

Consider two graphs $\ccalG_{P} = (\ccalV_{P},\ccalE_{P})$ and $\ccalG_{Q} = (\ccalV_{Q},\ccalE_{Q})$ with $P$ and $Q$ nodes, respectively. Let the corresponding graph Laplacian matrices be $ \bbL_{P} \in \reals^{P \times P}$ and $ \bbL_{Q} \in \reals^{Q \times Q}$. Let the Cartesian product of two graphs $ \ccalG_{P} $ and $ \ccalG_{Q}$ be denoted by $\ccalG_{N} $ with $ |\ccalV_{N}| =  |\ccalV_{P}||\ccalV_{Q}|$ nodes, i.e., $PQ = N$. In other words, the graphs $\ccalG_{P}$ and $\ccalG_{Q}$ are the factors of $\ccalG_{N}$. Then the graph Laplacian matrix $\bbL_{N}$ can be expressed in terms of $\bbL_{P}$ and $\bbL_Q$ as 
\begin{equation}\label{eq:cartesian_product}
\begin{aligned}
\bbL_{N}  = \bbL_{P} \oplus \bbL_{Q} = \bbI_Q \otimes \bbL_{P}  +   \bbL_{Q} \otimes \bbI_P. 
\end{aligned}
\end{equation}  

Let us collect the set of graph signals $\{\bbx_{i} \}^T_{i=1}$ with $\bbx_i \in \reals^N$ defined on the product graph $\ccalG_N$ in an $N \times T$ matrix $\bbX = [\bbx_1, \bbx_2,...,\bbx_T]$. As each node in the product graph $\ccalG_N$ is related to a pair of vertices in its graph factors, we can reshape any product graph signal $\bbx_i$ as $\bbX_i \in \reals^{P \times Q}$, $i=1,2,\ldots,T$, such that $\bbx_i = {\rm vec}(\bbX_i)$.
This means that each product graph signal represents a multidomain graph signal where the columns and rows of $\bbX_i$ are graph signals associated to the graph factor $\ccalG_P$ and $\ccalG_Q$, respectively. 

In this work, we will assume that $\bbX$ is smooth with respect to (w.r.t.) the graph $\ccalG_{N}$. The amount of smoothness is quantified by the Laplacian quadratic form ${\rm tr}(\bbX^T \bbL_{N} \bbX)$, where  small values of ${\rm tr}(\bbX^T \bbL_{N} \bbX)$ imply that the data $\bbX$ is smooth on the graph $\ccalG_N$. 

\section{Task-cognizant product graph learning}\label{ProbelmStatement}

Suppose we are given the training data $\bbX \in \reals^{N \times T}$ defined on the product graph $\ccalG_{N}$, where each column of $\bbX$, i.e., $\bbx_i \in \reals^{N}$, represents an multidomain graph data $\bbX_i \in \reals^{P \times Q}$.
Assuming that the given data is smooth on the graph $\ccalG_N$ and the product graph $\ccalG_N$ can be factorized as the Cartesian product of two graphs $\ccalG_P$ and $\ccalG_Q$ as in \eqref{eq:cartesian_product} we are interested in estimating the graph Laplacian matrices $\bbL_P$ and $\bbL_Q$. To do so, we assume that $P$ and $Q$ are known.

Typically, we might not have access to the original graph data $\bbX$, but we might observe data related to $\bbX$. Let us call this data $\bbY \in \reals^{N \times T}$. For example, $\bbY$ could be a noisy or an incomplete version of $\bbX$.
Given $\bbY$, the joint estimation of $\bbX$, $\bbL_P$ and $\bbL_Q$, may be mathematically formulated as the following optimization problem

\begin{equation}\label{eq:Problem_modeling}
\begin{aligned}
& \underset{{\bbL_{P} \in \ccalL_P,\bbL_{Q} \in \ccalL_Q, \bbX}}{{\rm minimize}}
& & f(\bbX,\bbY) + \alpha {\rm tr}(\bbX^{T}(\bbL_{P}\oplus \bbL_{Q})\bbX)\\
& && \> + \beta_{1} \|\bbL_P \|_{F}^{2} + \beta_2 \|\bbL_Q \|_{F}^{2}, 
\end{aligned}
\end{equation} 
where $\ccalL_N := \{ \bbL \in \reals^{N \times N} | \bbL {\bf 1} = {\bf 0}, {\rm tr}(\bbL) = N, L_{ij} =  L_{ji}  \leq 0, i \neq j\}$ is the space of all the valid Laplacian matrices of size $N$. We use the trace equality constraint in this set to avoid a trivial solution.  The loss function $f(\bbX,\bbY)$ is appropriately chosen depending on the nature of the observed data. More importantly, we learn the product graph suitable for the task of minimizing $f(\bbX,\bbY)$. For instance, if the observed graphs signals are noisy as $\bby_i = \bbx_i + \bbn_i$ for $1 \leq i \leq T$ with the noise vector $\bbn_i$, then $f(\bbX,\bbY)$ is chosen as $\|\bbX-\bbY\|_F^2$.
A smoothness promoting quadratic term is added to the objective function with a positive regularizer $\alpha$.
The squared Frobenius norm of the Laplacian matrices with tuning parameters $\beta_1$ and $\beta_2$ in the objective function controls the distribution of edge weights. 
By varying $\alpha$, $\beta_1$ and $\beta_2$, we can control the sparsity (i.e., the number of zeros) of $\bbL_P$ and $\bbL_Q$, while setting $\beta_1 = \beta_2 = 0$ gives the sparsest graph.

\section{Solver by adapting existing works}\label{ExistingWorks}

Ignoring the structure of $\ccalG_{N}$ that it can be factorized as $\ccalG_P$ and $\ccalG_Q$, and given $\bbX$ (or its noisy version), one can learn the graph Laplacian matrix $\bbL_N$ using any one of the methods in~\cite{learnDong,kalofolias2016learn,chepuri2016learning,kalofolias2017learning}. In this section, we will develop a solver for the optimization problem \eqref{eq:Problem_modeling} based on the existing method in~\cite{learnDong}.

There exists several numerical approaches for factorizing product graphs, i.e., to find the graph factors $\bbL_P$ and  $\bbL_Q$ from $\bbL_{N}$~\cite{Hammack_HandBookofGraphProducts,Imrich:2008:TGT:1608956}.
One of the simplest and efficient ways to obtain the graph factors is by solving the convex optimization problem
\begin{equation}\label{eq:Cartesian_Factorization}
\underset{{\bbL_P \in \ccalL_P,\bbL_{Q} \in \ccalL_Q}}{{\rm minimize}} \| \bbL_N - \bbL_P \oplus \bbL_Q\|^2_F.
\end{equation}
The above problem can be solved using any one of the off-the-shelf convex optimization toolboxes. However, notice that this is a two-step approach, which requires computating a size-$N$ Laplacian matrix in the first step using~\cite{learnDong}, for instance. The first step is computationally much expensive as the number of nodes in the product graph $\ccalG_{N}$ increases exponentially with the increase in the number of nodes in either of the factor graphs $\ccalG_P$ and $\ccalG_Q$. In other words, finding the product graph Laplacian matrix and then factorizing it is computationally expensive and sometimes infeasible because of the huge memory requirement for storing the data. Therefore, in the next section, we propose a computationally efficient solution.

\section{Proposed solver}\label{ProposedSolver}
In this section, instead of a two-step approach, as discussed in Section~\ref{ExistingWorks}, we provide a one-step solution for estimating the Laplacian matrices of the factors of the product graph by leveraging the properties of the Cartesian graph product and the structure of the data.

Assuming that the training data is noise free and complete, that is, $\bbY = \bbX$, the quadratic smoothness promoting term ${\rm tr}(\bbX^T (\bbL_P \oplus \bbL_Q) \bbX)$ can be written as 
\begin{equation}\label{eq:traceExpansion}
{\rm tr}(\bbX^T (\bbL_P \oplus \bbL_Q) \bbX)= \sum_{i=1}^{T} \bbx_i^T(\bbL_P \oplus \bbL_Q)\bbx_i.
\end{equation}
Using the property of the Cartesian product that for any given matrices $\bbA$, $\bbB$, $\bbC$ and $\bbX$ of appropriate dimensions, the equation $\bbA\bbX + \bbX\bbB = \bbC$ is equivalent to $(\bbA \oplus \bbB^T) \rm vec(\bbX) = \rm vec(\bbC)$. Using the property that ${\rm tr}(\bbA^T\bbX) = {\rm vec}(\bbA)^T {\rm vec}(\bbX)$, \eqref{eq:traceExpansion} simplifies to
\begin{equation}\label{eq:traceExpansion3}
\begin{aligned}
& \sum_{i=1}^{T} \bbx_i^T(\bbL_P \oplus \bbL_Q)\bbx_i &&=\sum_{i=1}^{T}{\rm vec}(\bbX_i)^T {\rm vec}(\bbL_P \bbX_i + \bbX_i \bbL_Q)\\ \nonumber
&&&=\sum_{i=1}^{T}{\rm tr}(\bbX_i^T \bbL_P \bbX_i) + {\rm tr}(\bbX_i \bbL_Q \bbX_i^T). \nonumber
\end{aligned}
\end{equation}
This means that the amount of smoothness of the multidomain signal $\bbx_i$ w.r.t. $\ccalG_N$ is equal to the sum of the amount of smoothness of the signals collected in the rows and columns of $\bbX_i$ w.r.t. $\ccalG_P$ and $\ccalG_Q$, respectively.

Therefore, with $\bbX$ available, solving \eqref{eq:Problem_modeling} is equivalent to solving 
\begin{equation}\label{eq:EquivalentFormulation}
\begin{aligned}
& \underset{{\bbL_{P} \in \ccalL_P,\bbL_{Q} \in \ccalL_Q}}{{\rm minimize}}
&&\alpha \sum_{i=1}^{T}[ {\rm tr}(\bbX_i^T \bbL_P \bbX_i) + {\rm tr}(\bbX_i \bbL_Q \bbX_i^T)] \\
& && \> + \beta_{1} \|\bbL_P \|_{F}^{2} + \beta_2 \|\bbL_Q \|_{F}^{2}  
\end{aligned}
\end{equation} 
with variables $\bbL_{P}$ and $\bbL_Q$. The optimization problem \eqref{eq:EquivalentFormulation} has a unique minimizer as it is convex in $\bbL_P$ and $\bbL_Q$. In fact, it can be expressed as a quadratic program (QP) with fewer variables as compared to~\eqref{eq:EquivalentFormulation} by exploiting the fact that $\bbL_P$ and $\bbL_Q$ are symmetric matrices. In essence, we need to solve for only the lower or upper triangular elements of $\bbL_P$ and $\bbL_Q$. 

Let the vectorized form of the lower triangular elements of $\bbL_P$ and $\bbL_Q$ be denoted by ${\rm vecl}(\bbL_P) \in \reals^{P(P+1)/2}$ and ${\rm vecl}(\bbL_Q) \in \reals^{Q(Q+1)/2}$, respectively. Furthermore, ${\rm vecl}(\bbL_P)$ and ${\rm vecl}(\bbL_Q)$, may be, respectively, related to ${\rm vec}(\bbL_P)$ and ${\rm vec}(\bbL_Q)$ as ${\rm vec}(\bbL_P) = \bbA {\rm vecl}(\bbL_P)$, ${\rm vec}(\bbL_Q) = \bbB {\rm vecl}(\bbL_Q)$ using matrices $\bbA$ and $\bbB$ of appropriate dimensions.
Now, using the fact that ${\rm tr}(\bbX^T \bbL_P \bbX) = {\rm vec}(\bbX\bbX^T)^T {\rm vec}(\bbL_P)$ and 
${\rm tr}(\bbX \bbL_Q \bbX^T) = {\rm vec}(\bbX^T\bbX)^T {\rm vec}(\bbL_Q)$ and the properties of Frobenius norm, we may rewrite \eqref{eq:EquivalentFormulation} as
\begin{equation}\label{eq:QpFormulation}
 \underset{{\bbz \in \reals^{K}}}{{\rm minimize}} \quad
 \dfrac{1}{2} \bbz^T \bbP \bbz + \bbq^T\bbz 
 \quad \text{subject to } \quad \bbC\bbz = \bbd, \,\, \bbz \geq \bb0,
\end{equation}
where $\bbz = \left[{\rm vecl}^T(\bbL_P), {\rm vecl}^T(\bbL_Q)\right]^T$ is the optimization variable of length $K = 0.5(P^2+Q^2+ P+ Q)$.  Here, $\bbP = {\rm diag}[2\beta_1 \bbA^T \bbA,$ $2\beta_2 \bbB^T \bbB]$ is the diagonal matrix of size $K$ and $\bbq =\alpha \sum_{i=1}^T \bbq_i$ with $\bbq_i = [{\rm vec}(\bbX_i\bbX^T_i)^T \bbA, {\rm vec}(\bbX^T_i\bbX_i)^T \bbB]^T \in \reals^{K}$. The matrix  $\bbC$ and the vector $\bbd$ are defined appropriately to represent the trace equality constraints (in the constraint sets) in \eqref{eq:EquivalentFormulation}. The problem \eqref{eq:QpFormulation} is a QP in its standard form and can be solved optimally using any one of the off-the-shelf solvers such as CVX~\cite{cvx}.
%

To obtain the graph factors $\bbL_P$ and $\bbL_Q$ using the solver based on the existing methods as described in Section~\ref{ExistingWorks} requires solving a QP with $N(N+1)/2$ variables for $\bbL_N$~\cite{learnDong}, and subsequently, factorizing $\bbL_N$ as $\bbL_P$ and $\bbL_Q$ as in \eqref{eq:Cartesian_Factorization}, which requires solving one more QP with $K$ variables. In contrast, the proposed method requires solving only one QP with $K$ variables. Thus, the computation complexity of the proposed method is very less as compared to the solver based on the existing methods.

\begin{figure*}[!h]
    \centering
    \psfrag{celcius}{\footnotesize $^\circ$ C}
    \psfrag{5}{\tiny 5} \psfrag{6}{\tiny 6} \psfrag{7}{\tiny 7} \psfrag{8}{\tiny 8} \psfrag{9}{\tiny 9} \psfrag{10}{\tiny 10} \psfrag{11}{\tiny 11}
    \begin{subfigure}[]{0.3\textwidth}
        \includegraphics[width=\columnwidth]{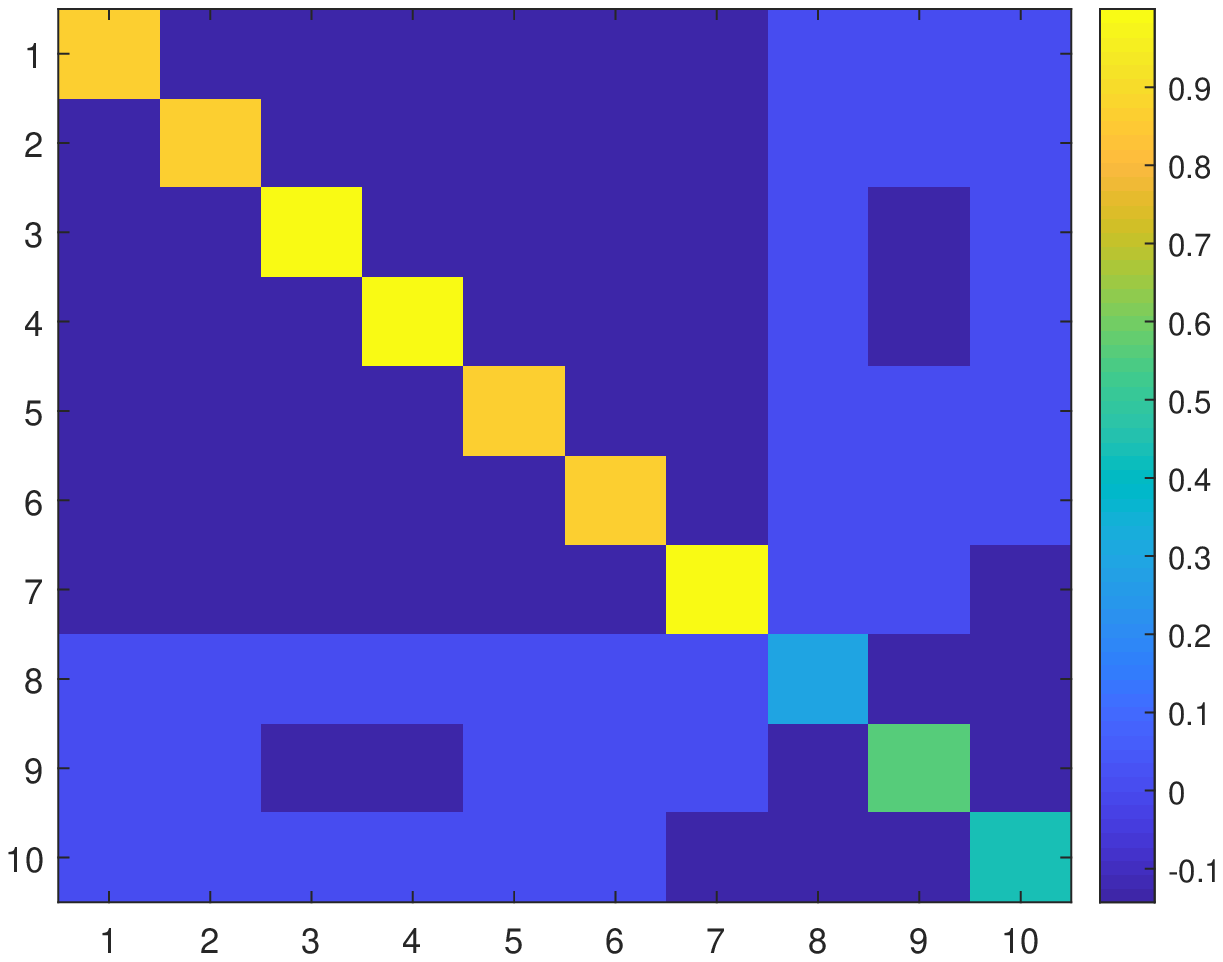}
        \caption{Ground truth: $\bbL_P$}
        \label{fig:Learned_Laplacian_P}
    \end{subfigure}
\begin{subfigure}[]{0.3\textwidth}
    \includegraphics[width=\columnwidth]{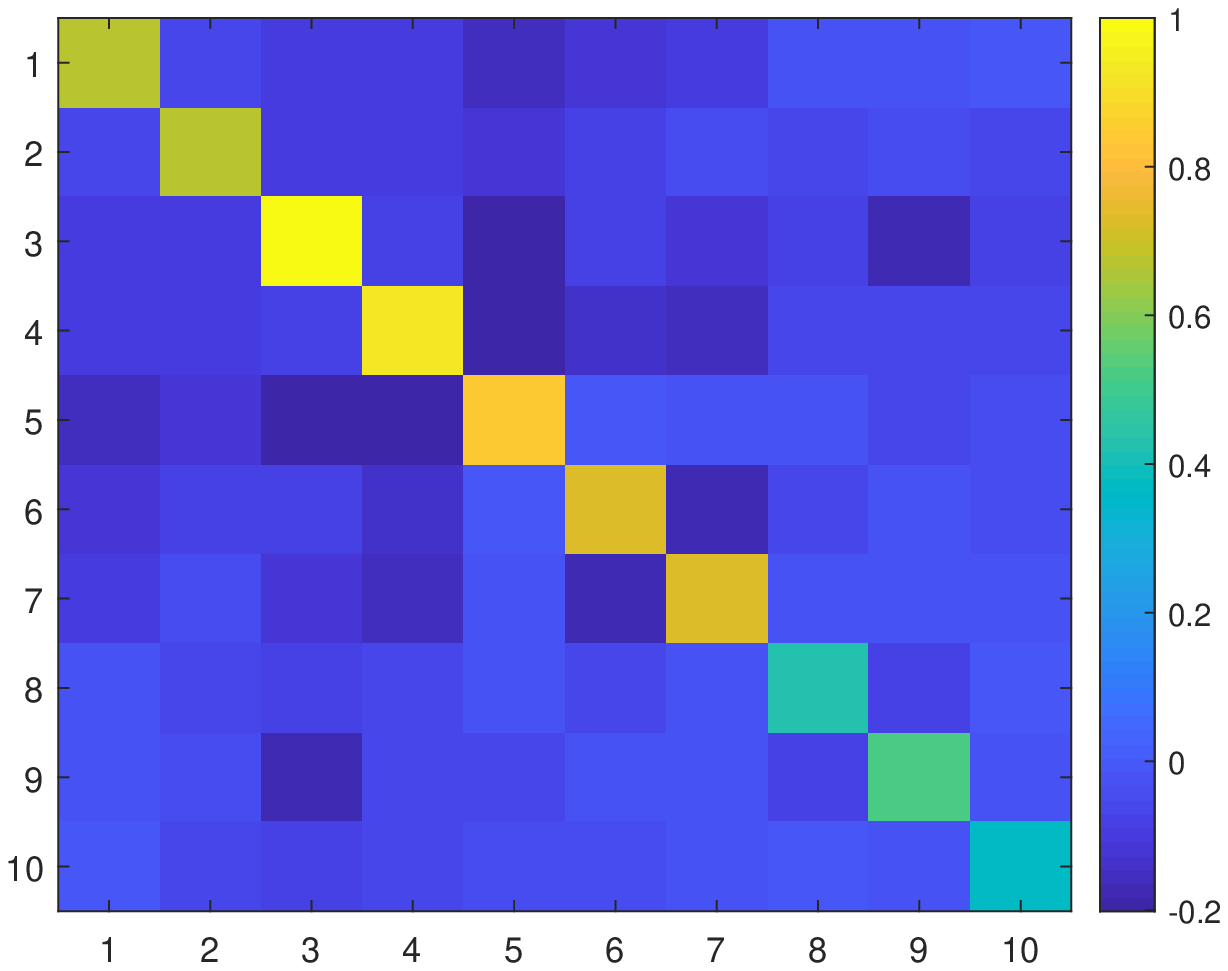}
    \caption{Learned: $\bbL_P$ (Solver 2)}
    \label{fig:Learned_Laplacian_Q}
\end{subfigure}
    \begin{subfigure}[]{0.3\textwidth}
        \includegraphics[width=\columnwidth]{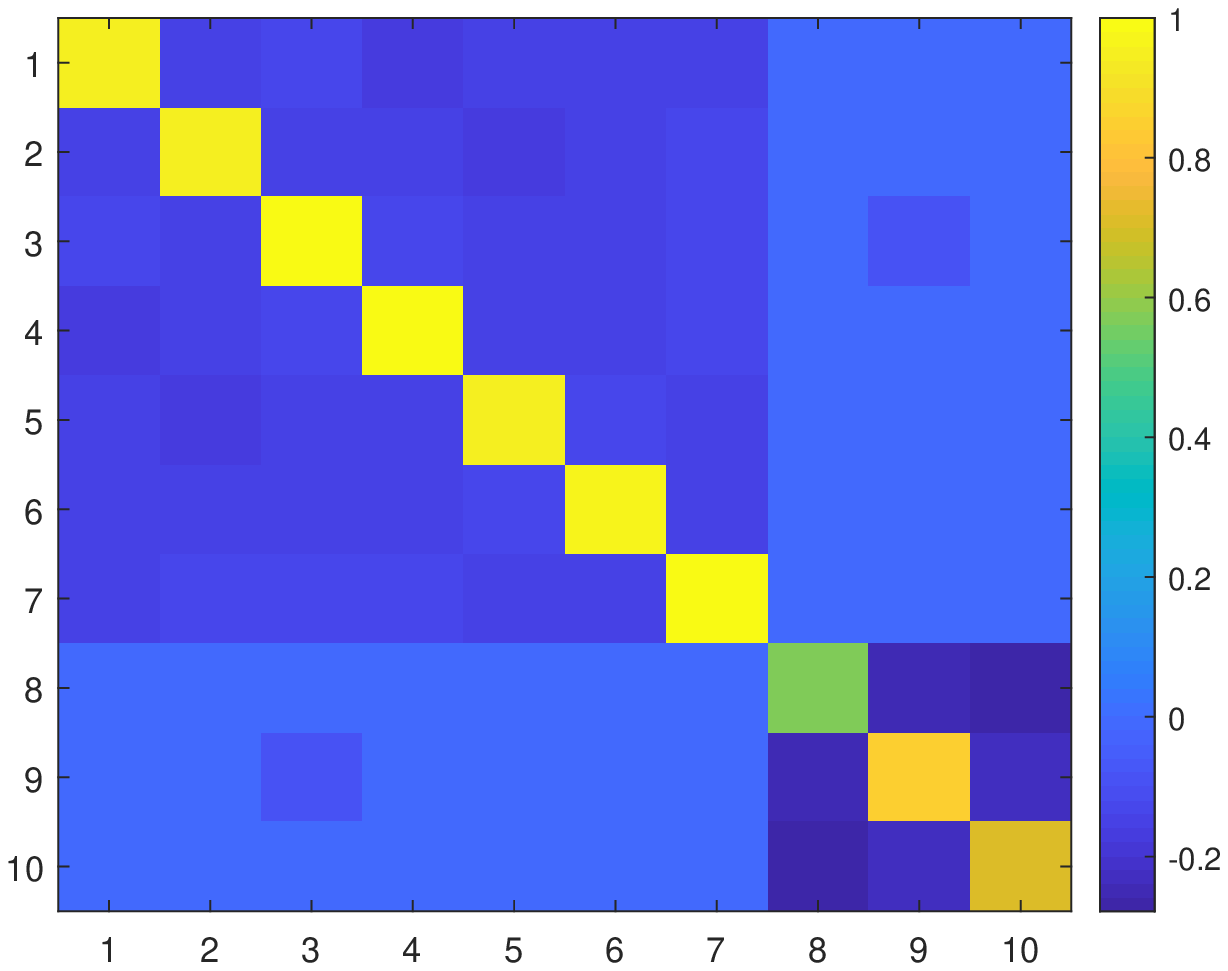}
        \caption{Learned: $\bbL_P$ (Solver 1)}
        \label{fig:Learned_Laplacian_P}
    \end{subfigure}%

    \begin{subfigure}[]{0.3\textwidth}
    \includegraphics[width=\columnwidth]{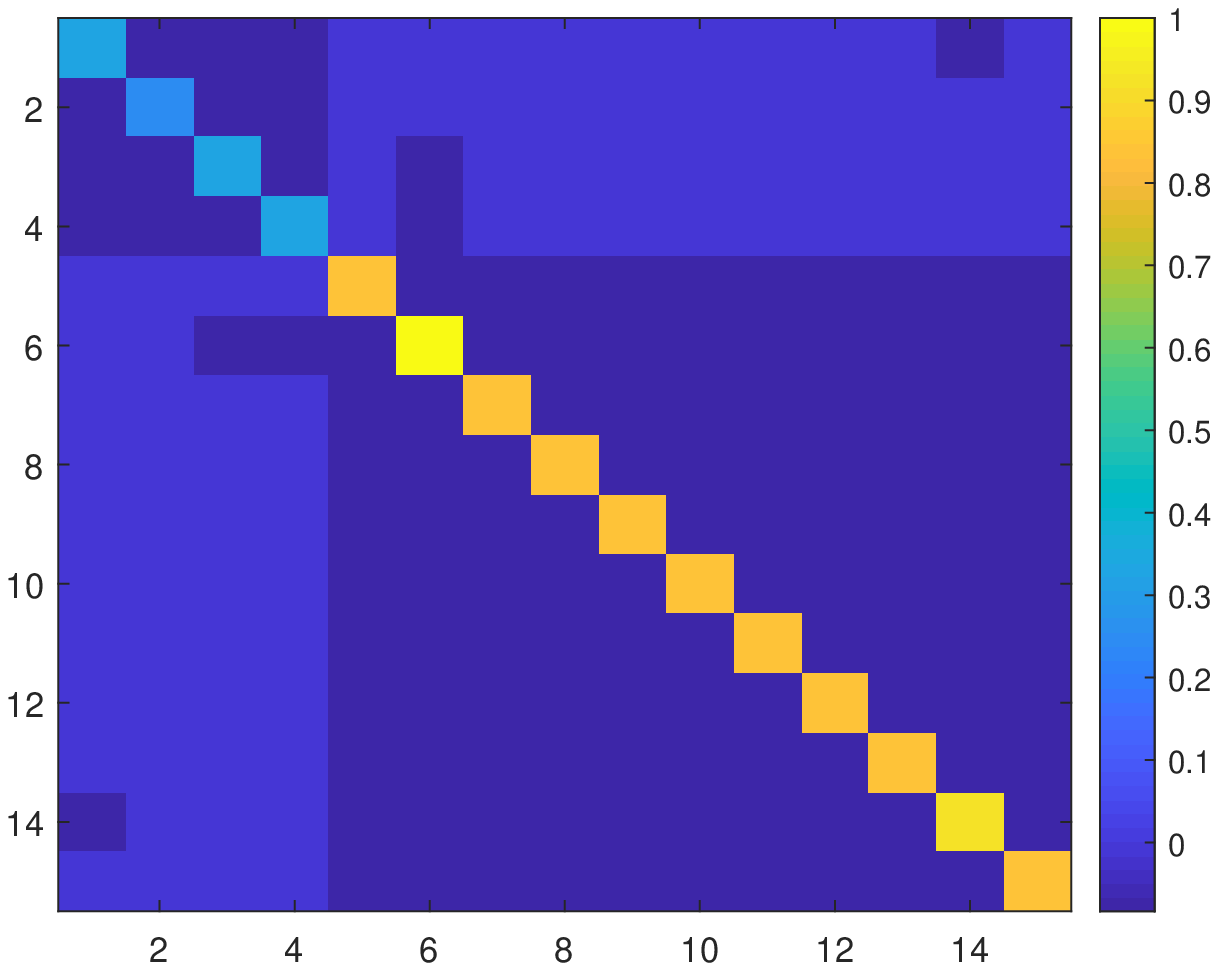}
    \caption{Ground truth: $\bbL_Q$}
    \label{fig:Learned_Laplacian_P}
\end{subfigure}
\begin{subfigure}[]{0.3\textwidth}
    \includegraphics[width=\columnwidth]{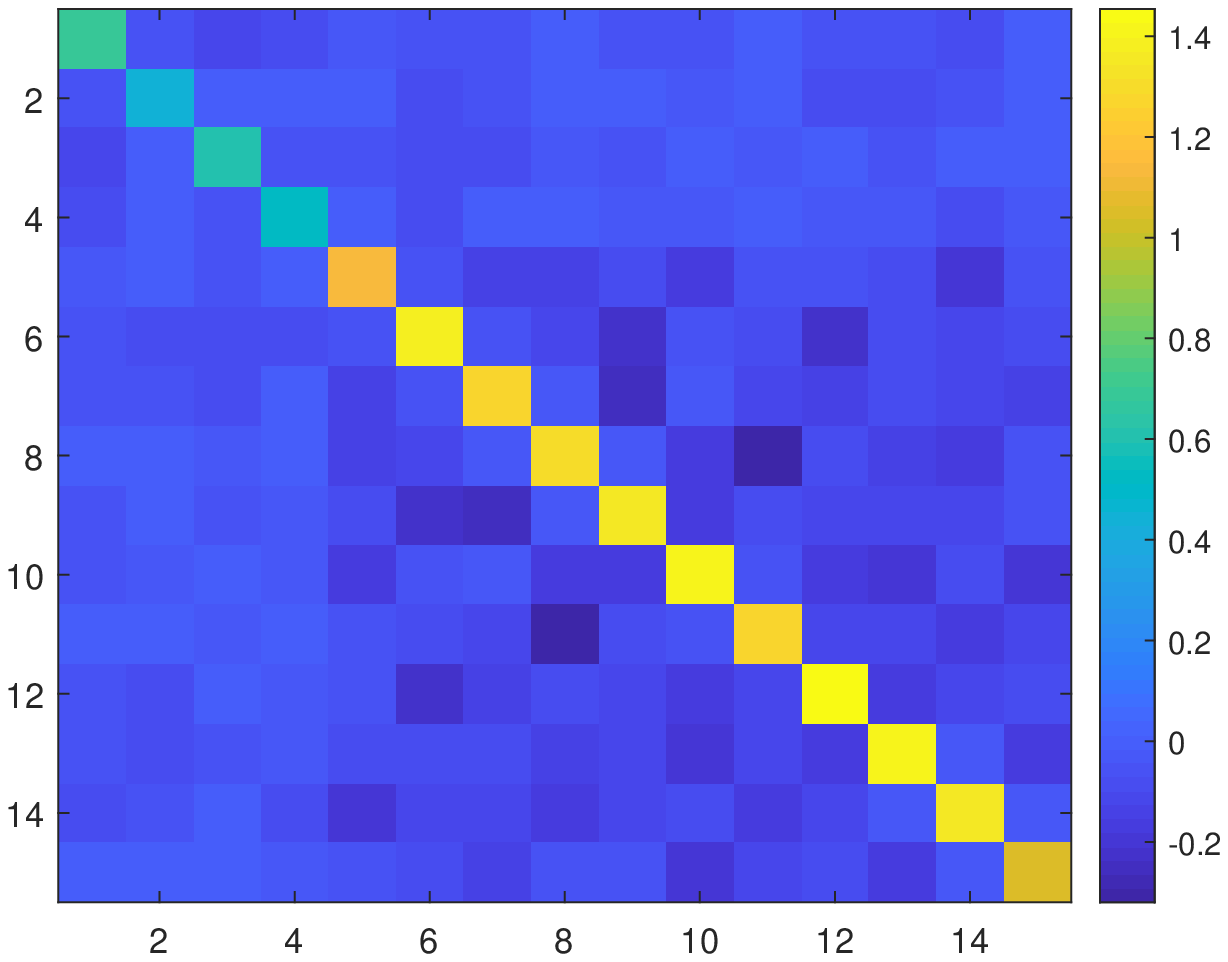}
    \caption{Learned: $\bbL_Q$ (Solver 2)}
    \label{fig:Learned_Laplacian_Q}
\end{subfigure}
\begin{subfigure}[]{0.3\textwidth}
    \includegraphics[width=\columnwidth]{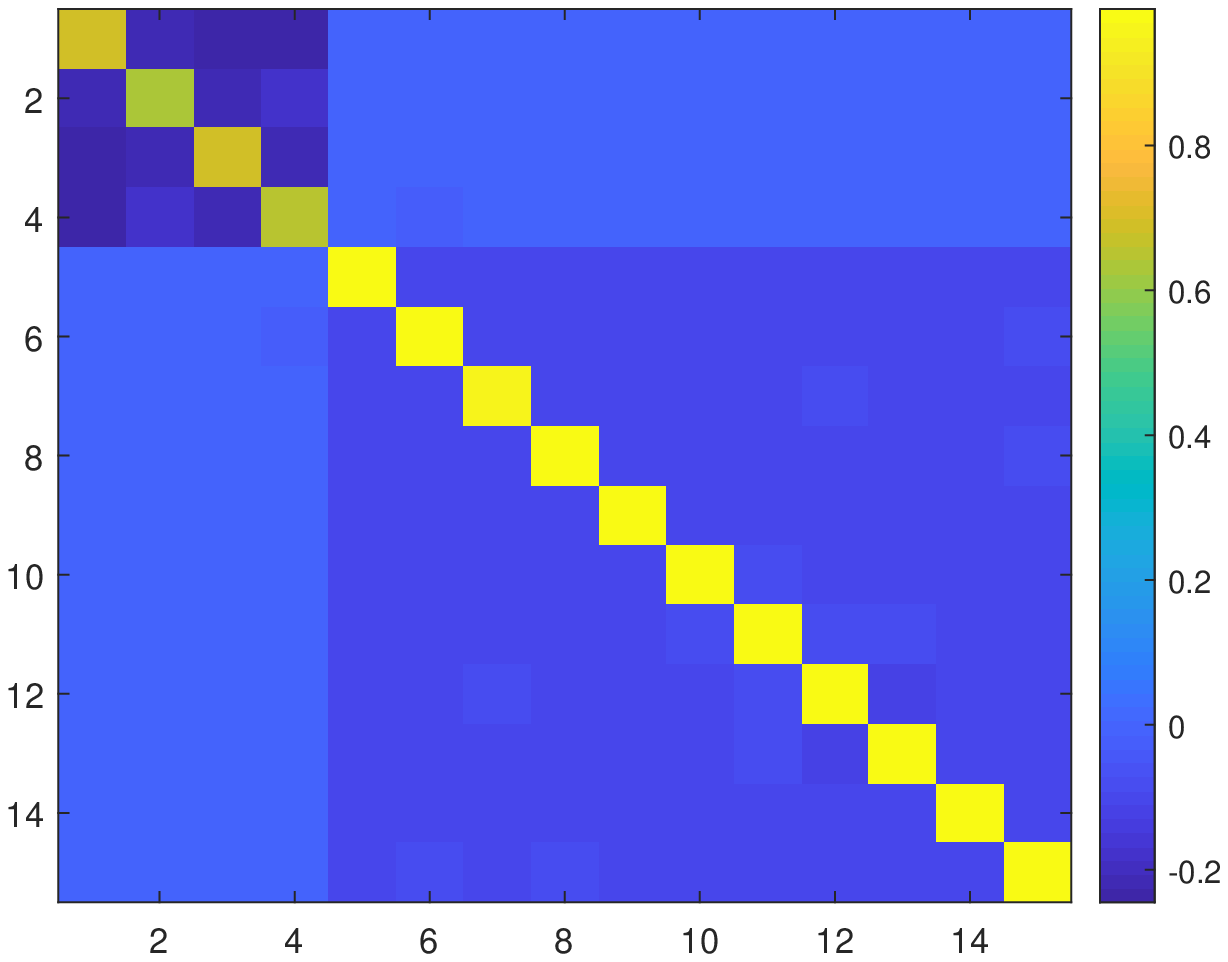}
    \caption{Learned: $\bbL_Q$ (Solver 1)}
    \label{fig:Learned_Laplacian_P}
\end{subfigure}%
\caption{\footnotesize{Product graph learning on synthetic data}}
\label{fig:Laplacians}
    \vskip-5mm
\end{figure*}

\begin{figure*}[!h]
    \centering
    \psfrag{celcius}{\footnotesize $^\circ$ C}
    \psfrag{5}{\tiny 5} \psfrag{6}{\tiny 6} \psfrag{7}{\tiny 7} \psfrag{8}{\tiny 8} \psfrag{9}{\tiny 9} \psfrag{10}{\tiny 10} \psfrag{11}{\tiny 11}
    \begin{subfigure}[t]{0.4\textwidth}
        \includegraphics[width=\columnwidth, height=2.5in]{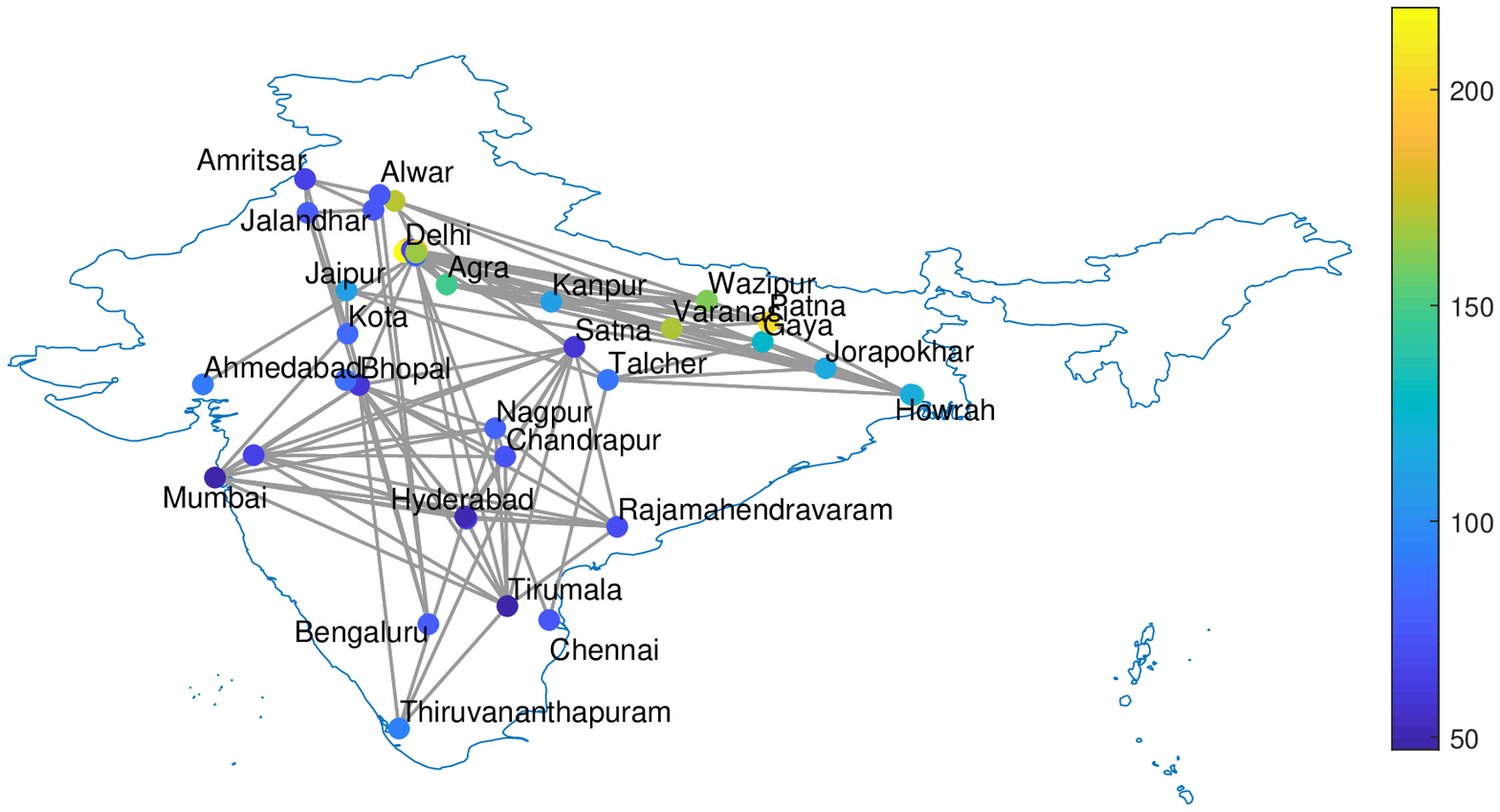}
        \caption{Graph factor $\ccalG_P$ }
        \label{fig:Graph_Laplacian_1}        
    \end{subfigure}%
~
\begin{subfigure}[t]{0.4\textwidth}
    \centering
    \includegraphics[width=\columnwidth]{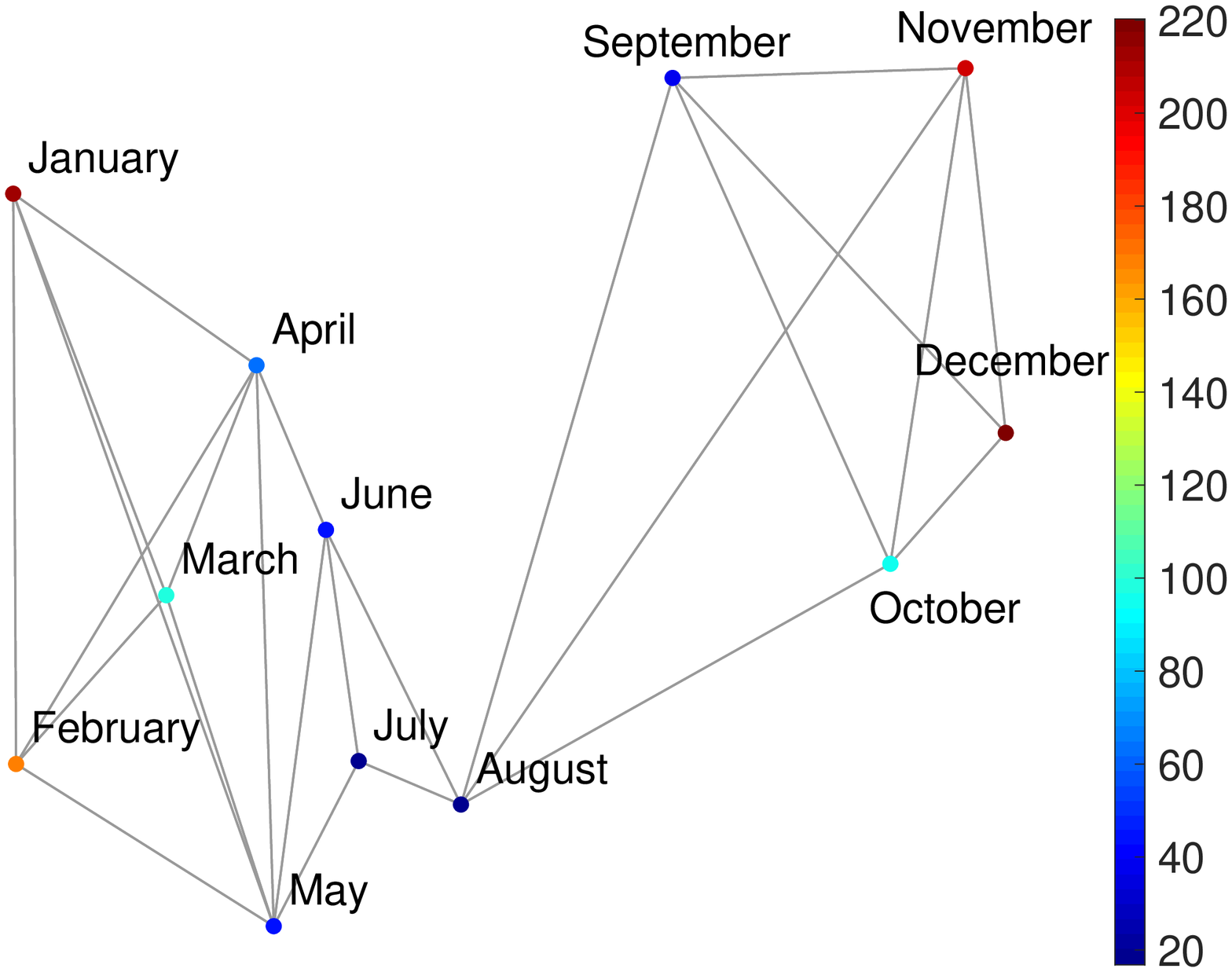}
    \caption{Graph factor $\ccalG_Q$}
    \label{fig:Graph_Laplacian_2}
\end{subfigure}     
    \caption{\footnotesize{\emph{Factor graph learning on air pollution data}: The colored dots indicates the ${\rm PM}_{2.5}$ values. (a) Graph factor $\ccalG_P$ represents the spatially distributed sensor graph. (b) Graph factor $\ccalG_Q$ is a graph capturing the temporal/seasonal variation of the $PM_{2.5}$ data across months.}}
    \label{fig:factor_graphs}
    \vskip-6mm
\end{figure*}
\section{Water-filling solution} \label{sec:waterfil}
In this section, we derive an explicit solution for the optimization problem \eqref{eq:QpFormulation} based on the well-known {\it water-filling} approach. By writing the KKT conditions and solving for $z_i$ leads to an explicit solution.

The Lagrangian function for \eqref{eq:QpFormulation} is given by
\begin{equation*}\label{eq:Lagrangian}
\ccalL(\bbz,\,\bblam,\,\bbmu) = \frac{1}{2} \bbz^T \bbP \bbz + \bbq^T \bbz + \bbmu^T(\bbd -\bbC\bbz) -\bblam^T \bbz,
\end{equation*}
where $\bblam$ and $\bbmu$ are the Lagrange multipliers corresponding to the inequality and equality constraints. Then the KKT conditions are given by
\begin{equation*}
\begin{aligned}
&P_{i,i}z_i^\star + q_i - \bbc_i^T \bbmu^\star - \lam_i^\star = 0, \\
\sum_{i=1}^{N} \bbc_i z_i^\star &= \bbd, \quad z_i^\star \geq 0,  \quad \lam_i^\star \geq 0, \quad \lam_i^\star z_i^\star = 0,
\end{aligned}
\end{equation*}
where $\bbc_i$ is the $i$th column of $\bbC$.
We can now solve the KKT conditions to find $z_i^\star$, $\bblam^\star$ and $\bbmu^\star$. To do so, we eliminate the slack variable $\lambda_i^\star$ and solve for $z_i^\star$. This results in 
\[
z_i^\star = {\rm max}\left\{0, P_{i,i}^{-1}(\bbc_i^T\bbmu^\star - q_i^\star)\right\}.
\]
To find $\bbmu^\star$, we may use $z_i^\star$ in the second KKT condition to obtain
$
\sum_{i=1}^{N} \bbc_i {\rm max}\{0, P_{i,i}^{-1}(\bbc_i^T\bbmu^\star - q_i^*)\}= \bbd.
$
Since this is a linear function in $\mu_i^\star$, we can compute $\mu_i^\star$ using a simple iterative method. This approach is similar to the water-filling algorithm with multiple water levels~\cite{SimpleWaterfill}. This method is computationally very cheap compared to solving a standard QP.
%

\section{Numerical results}
In this section, we will evaluate the performance of the proposed water-filling based solver (henceforth, referred to as Solver 1) and compare it with the solver based on the existing methods described in Section~\ref{ExistingWorks} (henceforth, referred to as Solver 2) on synthetic and real datasets.
\vspace*{-3mm}
\subsection{Results on synthetic data}

To evaluate the quantitative performance of the proposed method, we generate synthetic data on a known graph, which can be factorized. We will use those graph factors as a reference (i.e., ground truth) and compare the estimated graph factors with the ground truth in terms of F-measure, which is a measure of the percentage of correctly recovered edges~\cite{learnDong}. 
Specifically, we generate a graph with $N = 150$ nodes, which is formed by the Cartesian product of two community graphs with $P= 10$ and $Q = 15$ nodes, respectively. 

We generate $T = 50$ signals $\{\bbx_i\}^{50}_{i=1}$ that are smooth w.r.t. $\ccalG_{N}$ by using the factor analysis model described in \cite{learnDong}. Given $\bbX$, we learn the graph factors using Solver 1 (using the water-filling method) and Solver 2 (wherein we learn the complete graph Laplacian matrix of $N=150$ nodes as in \cite{learnDong} and then factorize it by solving \eqref{eq:Cartesian_Factorization}).

%
In Table~\ref{tab:my-table}, we show the best (in terms of the tuning parameters) F-measure scores obtained by Solver 1 and Solver 2.  The F-measure of Solver 1 is close to $0.98$, which means that the proposed method learns the graph factors close to the ground truth.
In Fig. \ref{fig:Laplacians}, we plot the Laplacian matrices of the ground truth, learned Laplacian matrices from Solver 1 and Solver 2. We see that the graph factors estimated from Solver 1 are more consistent with the ground truth than the factors estimated using Solver 2. 
Moreover, we stress the fact that the computational complexity of the proposed water-filling method is very less as compared to Solver 2.
\begin{table}[!t]
    \centering
    \begin{tabular}{|l|l|l|l|}
        \hline
    \textbf{Method}    & $\bbL_{P}$ & $\bbL_{Q}$ & $\bbL_{N}$  \\ \hline
        \textbf{Solver 1}& 0.9615 & 0.9841 & 0.9755 \\ \hline
        \textbf{Solver 2}& 0.7556 & 0.7842  & 0.7612 \\ \hline
    \end{tabular}
    \caption{F-measure of the proposed  solver (Solver 1) and the solver based on the existing methods (Solver 2).}
    \label{tab:my-table}
    \vspace*{-5mm}
\end{table}

\subsection{Joint matrix completion and learning graph factors}
We now test the performance on real data. For this purpose, we use the ${\rm PM}_{2.5}$ data collected over $40$ air quality monitoring stations in different locations in India for each day of the year $2018$~\cite{aqi}. However, there are many missing entries in this dataset. Given this multidomain data that has spatial and temporal dimensions, the aim is to learn the graph factors that best explain the data.
More precisely, we aim to learn the graph factors that capture the relationships between spatially distributed sensors and a graph that captures the seasonal variations of the ${\rm PM}_{2.5}$ data. Since the dataset has missing entries, we impute the missing entries using a graph Laplacian regularized nuclear norm minimization~\cite{kalofolias2014matrix}.
That is, we use $f(\bbX,\bbY) := \sum_{i=1}^{T} \|\ccalA (\bbX_i -\bbY_i)\|^2_F + \|\bbX_i\|_*$, where $\ccalA$ denotes the known observation mask that selects the available entries and $\|\cdot\|_*$ denotes the nuclear norm. As the problem \eqref{eq:Problem_modeling} is not convex in $\{\bbX, \bbL_P, \bbL_Q\}$ due to the coupling between the variables in the smoothness promoting terms, we use alternating minimization method. Specifically, in an alternating manner, we solve for $\{\bbL_P, \bbL_Q\}$ by fixing $\bbX$ using the solver described in Section~\ref{sec:waterfil}, and then solve for $\bbX$ by fixing $\{\bbL_P, \bbL_Q\}$ as
\begin{equation*}\label{eq:MatrixCompletion}
\underset{\{\bbX_i\}_{i=1}^T}{\text{minimize}} \,\,
f(\bbX,\bbY)  \,\, 
 + \,\, \alpha \sum_{i=1}^{T}  [{\rm \tr}(\bbX_i^T \bbL_P \bbX_i) + {\rm \tr}(\bbX_i\bbL_Q\bbX_i^T)],
\end{equation*}
where recall that ${\rm vec}(\bbX_i)$ forms the $i$th column of $\bbX$.

The learned graph factors that represent the multidomain graph signals are shown in  Fig.~\ref{fig:factor_graphs}. Fig.~\ref{fig:Graph_Laplacian_1} shows the graph factor that encodes the relationships between the spatially distributed air quality monitoring stations. Each colored dot indicates the concentration of ${\rm PM}_{2.5}$ on a random day across different locations. We can see from Fig.~\ref{fig:Graph_Laplacian_1} that the graph connections are not necessarily related to the geographical proximity of the cities, but based on the similarity of the ${\rm PM}_{2.5}$ values.

The concentration of ${\rm PM}_{2.5}$ is relatively lesser during the summer/monsoon months (i.e., June, July, and August) as compared to the other months. The graph capturing these temporal variations in the ${\rm PM}_{2.5}$ data is shown in Fig.~\ref{fig:Graph_Laplacian_2}. 
Each color dot indicates the concentration of ${\rm PM}_{2.5}$ averaged over each month at a random location. We can see in Fig.~\ref{fig:Graph_Laplacian_2} that months with similar average concentration of ${\rm PM}_{2.5}$ are connected. Moreover, the months having low ${\rm PM}_{2.5}$ values (during the monsoon months) are clustered together and the remaining months are clustered into two groups based on their occurrence before and after the monsoon. This shows that the obtained time graph has clusters that capture the seasonal variations hence confirms the quality of the learned graph.

\section{Conclusions}
\vspace*{-5mm}
We developed a framework for learning graphs that can be factorized as the Cartesian product of two smaller graphs. The estimation of the Laplacian matrices of the factor graphs is posed as a convex optimization problem. The proposed solver is computationally efficient and has an explicit solution based on the water-filling method. The performance of the proposed method is significantly better than other intuitive ways to solve the problem. We present numerical experiments based on air pollution data collected across different locations in India. Since this dataset has missing entries, we developed a task-cognizant learning method to solve the joint matrix completion and product graph learning problem.

\pagebreak

\bibliographystyle{IEEEtran}
\bibliography{IEEEabrv,sample}

\end{document}